\documentclass[manuscript, noacm]{acmart}

\usepackage{graphicx} 
\usepackage{xcolor}

\usepackage{paralist}
\usepackage{hyperref}

\hypersetup{
 colorlinks=true,
 citecolor=gray,
 linkcolor=blue,
 urlcolor=blue
}
\usepackage{multirow}
\usepackage{array}
\newtheorem{definition}{Definition}
\setcopyright{none}
\renewcommand\footnotetextcopyrightpermission[1]{} 

\title[Revolutionizing Datacenter Networks]{Revolutionizing Datacenter Networks via Reconfigurable Topologies}
 
\author{Chen Avin}
\affiliation{is a Professor at Ben-Gurion University of the Negev, Beersheva\country{Israel}}

\author{Stefan Schmid}
\affiliation{is a Professor at TU Berlin, Berlin\country{Germany}}

\begin{document}

\begin{abstract}
With the popularity of cloud computing and data-intensive applications such as machine learning, datacenter networks have become a critical infrastructure for our digital society. Given the explosive growth of datacenter traffic and the slowdown of Moore’s law, significant efforts have been made to improve datacenter network performance over the last decade. A particularly innovative solution is reconfigurable datacenter networks (RDCNs): datacenter networks whose topologies dynamically change over time, in either a demand-oblivious or a demand-aware manner. Such dynamic topologies are enabled by recent optical switching technologies and stand in stark contrast to state-of-the-art datacenter network topologies, which are fixed and oblivious to the actual traffic demand. In particular, reconfigurable demand-aware and “self-adjusting” datacenter networks are motivated empirically by the significant spatial and temporal structures observed in datacenter communication traffic. This paper presents an overview of reconfigurable datacenter networks. In particular, we discuss the motivation for such reconfigurable architectures, review the technological enablers, and present a taxonomy that classifies the design space into two dimensions: static vs. dynamic and demand-oblivious vs. demand-aware. We further present a formal model and discuss related research challenges.
Our article comes with complementary video interviews in which three leading experts, Manya Ghobadi, Amin Vahdat, and George Papen, share with us their perspectives on reconfigurable datacenter networks.
\end{abstract}

\maketitle

\section*{Key Insights}

\begin{itemize}

\item {\bf Datacenter networks have become a critical infrastructure} for our digital society, serving explosively growing communication traffic. 
\item {\bf Reconfigurable datacenter networks (RDCNs)} which can adapt their topology dynamically, based on innovative {\bf optical switching technologies}, bear the potential to improve datacenter network performance, 
and to simplify datacenter planning and operations. 
\item Demand-aware dynamic topologies are particularly interesting because of the {\bf significant spatial and temporal structures} observed in real-world traffic, e.g., related to distributed machine learning. 
\item The study of RDCNs and self-adjusting networks raises many {\bf novel technological and research challenges} related to their design, control, and performance. 
\end{itemize}

\section{Introduction}

Datacenter networks have become critical infrastructure for our digital society. Indeed, the performance of many distributed systems and cloud applications, e.g., those related to distributed machine learning, batch processing, scale-out databases, or streaming, critically depends on the throughput capacity of the underlying network topology.

Fortunately, thus far, Moore’s law for networking – which states that electrical switches double in bandwidth every two years at the same power and cost – has allowed datacenter operators to scale up their networks across generations. However, this impressive free scaling of switches now lags behind the doubling of cloud traffic, which occurs roughly every year [39]. It is also expected that Moore’s law will slow further in the coming years [9]. Accordingly, significant efforts are being made to improve the throughput of datacenter networks, and leading companies such as Google, Microsoft, and Facebook are competing to improve the operations of their networks~\cite{singh2015jupiter,sirius,roy2015inside}.  

State-of-the-art datacenter network topologies have two common features: they are fixed and oblivious to the traffic demands they serve. In particular, current datacenter topologies are optimized for uniform, “all-to-all” traffic. This stands in stark contrast to the actual traffic patterns in datacenters, which often feature distinct spatial and temporal structures, i.e., they are skewed and bursty~\cite{kandula2009nature,roy2015inside}.
Thus, a particularly intriguing and novel approach to further improve datacenter performance is to leverage a new degree of freedom, related to the physical network topology. 

Emerging optical technologies enable reconfigurable datacenter networks (RDCNs): networks whose topologies can dynamically change to accommodate, or even adapt to and exploit, the structures in their communication traffic in a self-adjusting manner. Hence, RDCNs introduce resource allocation flexibility in the physical layer. This is also known as topology engineering: the ability to control the dynamic network topology. More specifically,
such dynamic datacenter topologies do not require any rewiring, but rather are enabled by optical circuit switches (OCSs) and similar technologies.
Depending on the specific technology used, the topological interconnects between racks can be changed within milliseconds, microseconds, or, recently, even nanoseconds.

However, there is still no consensus in the community on which RDCN design is optimal. In particular, while static network topologies are currently well understood and fat-tree topologies \cite{clos} have emerged as the de-facto topologies for datacenters, we still lack fundamental models and metrics for rigorously studying and comparing different RDCN designs.

In this article, we provide an overview of RDCNs. We review the motivation for RDCNs and discuss their technological enablers. We then present a taxonomy of the RDCN design space, and propose a formal model to raise interest in the topic in the research community. We conclude by discussing challenges ahead, also based on expert interviews. 

\section{Empirical Motivation: Structure in Application's Traffic}\label{sec:empirical}

\begin{quotation}
    \textit{“Think about a machine learning training job, say, training a Chat GPT model. It takes months to train this model, but the traffic matrix is beautifully predictable and periodic, which makes it very suitable to think about whether or not we could adjust the topology according to the traffic.”} 
    \par \hfill--Manya Gobhadi (MIT), Interviews, Section \ref{sec:Conclusion}
\end{quotation}

Datacenter traffic is growing explosively in volume \cite{poutievski2022jupiter}. Moreover, it features distinct structures: packet traces from real-world applications are far from arbitrary or random. Measurement studies~\cite{kandula2009nature,roy2015inside} have revealed that the traffic patterns in datacenters feature considerable locality, both in space and in time: they are skewed and bursty~\cite{sigmetrics20complexity}, and communication clusters are  relatively stable over time~\cite{foerster2023analyzing,roy2015inside}. Kandula et al.~\cite{kandula2009nature} showed that much of the traffic volume in production Microsoft datacenters could be explained by simple patterns. Facebook~\cite{roy2015inside} reported that in their datacenters, many flows are long- lived, and while hosts (servers) often communicate with hundreds of hosts and racks concurrently, the majority of traffic is often destined for only (a few) tens of racks.

\begin{figure}[t]
  \begin{centering}
  \begin{tabular}{>{\centering}m{0.45\textwidth} >{\centering}m{0.450\textwidth}}
     \multicolumn{2}{c}{\includegraphics[width=0.90\textwidth]{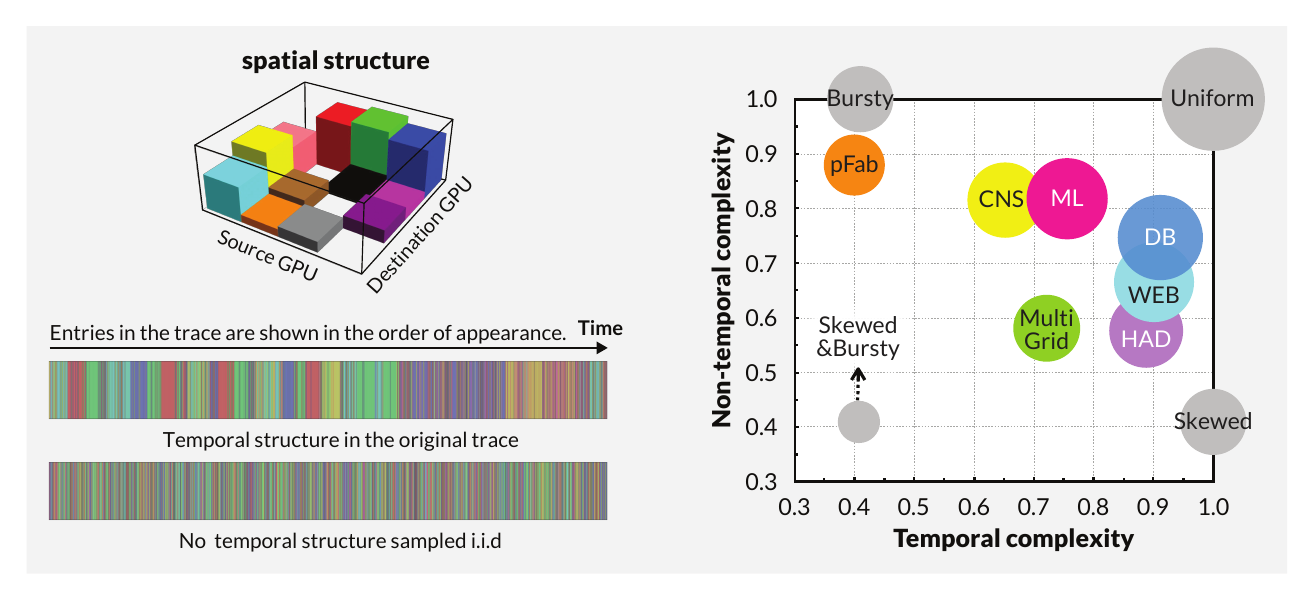}}\\
      (a) A real ML trace & (b) Complexity map
  \end{tabular}
        \caption{(a) Visualization of the spatial and temporal structures 
        of a simple packet trace (middle) from a machine learning (ML) application, a popular convolutional neural network training job, on four GPUs. 
        (top) the demand matrix. (bottom) an i.i.d trace with the same distribution.
        (b) A complexity map of seven real traces (colored circles) and four reference points placed at the corners of the map. 
        }
    \label{fig:locality}
  \end{centering}
\end{figure}

As a small-scale demonstrative example, consider Figure~\ref{fig:locality} (a) (middle), which shows a GPU-to-GPU trace that was generated by a neural network training job. This trace features both temporal and spatial locality (structure). The trace histogram, more specifically its demand matrix (top), is skewed, exhibiting a spatial structure. Moreover, compared to a uniform trace (bottom), the traffic is bursty in the time dimension, featuring a temporal structure.

The specific degree of traffic locality depends on the application~\cite{sigmetrics20complexity,roy2015inside}.  
A new methodology called \emph{complexity map}, provides an intuitive visualization of the strengths of the temporal and non-temporal structures in the traffic of different distributed applications~\cite{sigmetrics20complexity}. 
The complexity map measures the amounts of entropy in a trace along the space and time dimensions, and compares them to those of a completely unstructured (uniform) trace. Figure \ref{fig:locality} (b) shows such a complexity map for seven different real-world traces~\cite{roy2015inside}: three different Facebook datacenter traces, related to the web (WEB), database (DB), and Hadoop (HAD) applications; two high-performance computing benchmarks (CNS and, MultiGrid); a machine learning (ML) job; and a popular synthetic trace pFabric (pFab). For reference, it also shows a purely uniform traffic pattern, which is mapped to the upper right; a pattern that features only a temporal structure but no spatial structure, on the upper left; and a pattern that features only a spatial structure, on the lower right. We can see that the seven real-world traces indeed have different amounts of structure, and none of these traffic patterns is uniform.

The expanding variety of cloud applications not only creates additional traffic patterns that come in many different flavors, but also introduces different performance requirements. For instance, batch processing applications such as Hadoop exhibit an all-to-all traffic pattern that requires very high throughput; short query traffic has a skewed pattern and requires very low latency; and all-reduce operations in ML applications exhibit a ring or treelike traffic pattern, with a small number of elephant flows whose completion times critically affect the overall application performance~\cite{topoopt}. Traffic patterns can also change over time,; for example, one training iteration typically contains forward, backward, and all-reduce phases, resulting in periodic (and fairly predictable) demand~\cite{xiao2018gandiva}.

The main takeaway is that while datacenter communication traffic comes in different flavors, each still features some structure and therefore has low complexity. In turn, this structure provides opportunities for researchers to optimize the network topology accordingly.

\section{Traditional Datacenter Network Topology Design}\label{sec:traditional}

The most essential property of traditional datacenter networks is that they use static wired topologies. As datacenter traffic patterns evolve over time, these static topologies are designed to work well for \emph{any} communication input; thus, they are \emph{oblivious} to the real traffic, in the sense that they are optimized only for the worst-case scenario. Accordingly, common datacenter network topologies are regular and symmetric, with many redundant short paths to support high capacity and availability. Hosts (or other compute infrastructures, such as GPUs) in datacenters are usually mounted in racks, each of which is connected to the datacenter network via a top-of-rack switch (ToR).

Traditional datacenter networks are built from electrical switches connected by high-speed optical fibers. 
Most deployed datacenter networks rely on fat-tree topologies, which are based on Clos networks~\cite{clos}, and come in different flavors~\cite{clos,singh2015jupiter,roy2015inside}.
In contrast to other communication networks, datacenter networks typically operate under a single administrative domain, which facilitates simpler control using, e.g., software-defined networking (SDN). Such a single domain can include a unified controller and network operating system (NOS), and allows for fast and fine-grained adaptation of the network layer.

\begin{figure}
    \begin{centering}
     \begin{tabular}{>{\centering}m{0.5\textwidth} >{\centering}m{0.5\textwidth}}
     \multicolumn{2}{c}{\includegraphics[width=\textwidth]{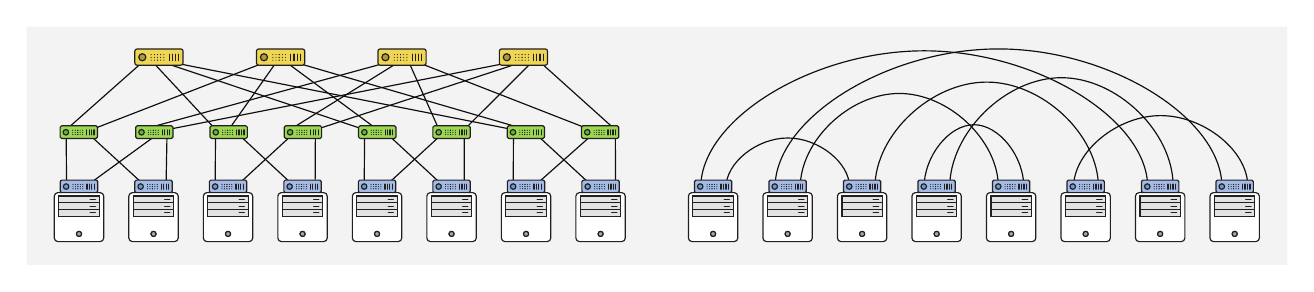}} \\
        (a) Fat-tree topology - bi-regular & (b) Expander topology - uni-regular
    \end{tabular}
    \caption{Examples of state-of-the-art static datacenter network topologies with eight racks, each containing two hosts. (a) A fat-tree, bi-regular topology with two types of switches, top-of-rack (ToR) switches (in blue) connecting to the hosts, and additional switches (in green) to increase throughout. (b) An expander-based, uni-regular topology, with only ToR switches. In both topologies, the diameter is four.}
    \label{fig:biregular}
    \end{centering}
\end{figure}

Existing datacenters typically aim to achieve a throughput metric called ``full bisection bandwidth'', which essentially means non-blocking traffic forwarding.
The bisection bandwidth refers to the smallest aggregate capacity of the links crossing the worst-case cut among all possible cuts that divide the topology graph into two halves. A network has full bisection bandwidth if its bisection bandwidth is equal to half of the total host traffic; topologies like fat-tree-based networks permit arbitrary application instance placement. 
To achieve this, fat-tree datacenter networks are typically based on a non-blocking multistage switching topology built out of smaller radix switch chips. In the resulting hierarchical designs, which are referred to as bi-regular topologies~\cite{uniregular2021throughput}, there are two types of switches: switches that connect to hosts (and from which traffic thus enters the network) and switches that connect only to other switches (which therefore merely forward traffic). Adding more switches of the former type can help increase the throughput between hosts. While such bi-regular datacenters provide the desired non-blocking architecture, this comes at the cost of requiring additional switches that are not connected to the hosts. Figure \ref{fig:biregular} (a) shows an example of a non-blocking, three-layer fat- tree topology that connects eight racks, where the two types of switches (each of radix four) are colored in blue and green.

Aiming to provide full bisection bandwidth at lower cost, another recent class of datacenter networks is based on \emph{uni-regular} topologies~\cite{uniregular2021throughput}, which can have lower installation and management overheads. Here, \emph{every} switch connects to a fixed number of hosts. 
Such networks also come in different flavors and their topologies are often based on expander graphs~\cite{xpander}. Figure \ref{fig:biregular} (b) shows an example of an expander-based, uni-regular, topology that connects eight racks using only ToR switches (of radix 4).

Alternative throughput metrics have also been presented in the literature on datacenter networks. The throughput under a given saturated\footnote{The sum of the rates in each row and column of
the demand matrix matches the switch capacities.} traffic matrix $T$ can be defined as the largest scaling factor $\theta(T)$ such that the topology can support the traffic matrix, without violating any link’s capacity constraint. 
The worst-case throughput of a topology among all traffic matrices is denoted by $\theta^*$. A topology can support any traffic demand if and only if $\theta^*$ is at least 1 (in this case, we say that the topology has full throughput). Because it can support any traffic demand, a full-throughput topology also permits arbitrary application instance placement by definition.
The bisection bandwidth and throughput metrics have recently been systematically compared, and it was found that depending on which metric is used, different conclusions may result, impacting cost and management complexity~\cite{uniregular2021throughput}.  In general, it is believed that throughput metrics better capture the capacity of both uni-regular and bi-regular topologies. 

Traditional datacenter network designs (whether bi-regular or uni-regular) all share the properties that they are fixed and oblivious to the real demand. 
This can come at a cost. In particular, static topologies inherently require multi-hop forwarding, which can introduce inefficiencies, especially for large flows: the more hops a flow
must traverse, the more network capacity is consumed. This overhead has been termed \emph{bandwidth tax}~\cite{opera} in the literature, and refers to the additional bytes that need to be sent along multiple hops (due to multi-hop routing) compared to direct, single-hop routing. 
In current fat-tree-based datacenter designs, due to the radix limitation of switches, the network diameter is usually 6 or 8 hops. 
Recently, it was shown that the \emph{expected route length} (i.e., bandwidth tax) and the throughput are inversely related in uni-regular designs: the higher the tax, the lower is the throughput \cite{uniregular2021throughput,griner2021cerberus,addanki2023mars}.

There are further limitations associated with the existing datacenter network designs. For example, while the fiber-optic links widely used in datacenters can meet high bandwidth demands, the optical–electrical–optical conversion necessary to connect such links to power-hungry electrical packet switches is costly and introduces inefficiencies and overhead~\cite{poutievski2022jupiter}.
Current datacenters are also often inflexible in that they are optimized for homogeneous hardware. For example, a fat-tree topology requires a spine layer with uniform support for the fastest devices that might connect to it. 
As hosts and storage devices are continuously evolving, without support for heterogeneous speeds, incremental deployments are expensive. Clearly, replacing the entire spine layer to support a faster speed would be impractical, involving hundreds of individual racks and thousands of fibers.

\begin{figure}
    \centering
    \begin{tabular}{>{\centering}m{0.62\textwidth} >{\centering}m{0.38\textwidth}}
     \multicolumn{2}{c}{\includegraphics[width=\textwidth]{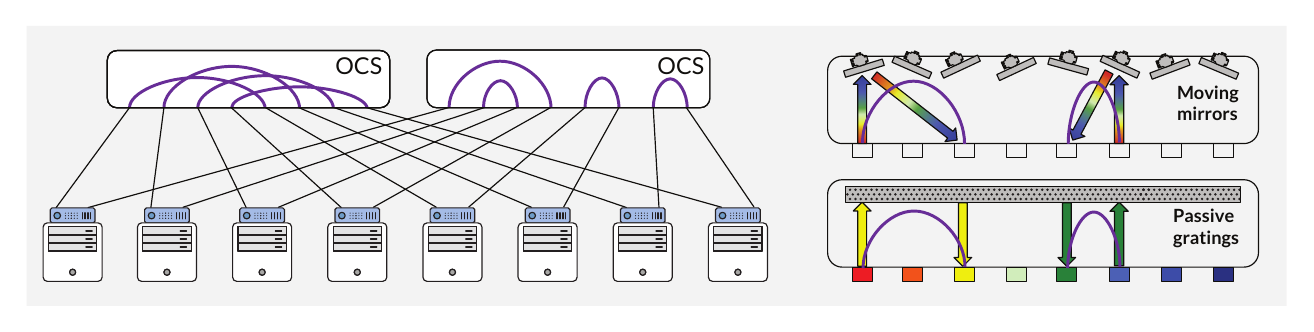}} \\
        
        (a) RDCN topology with two optical circuit switches (OCS) as spines & 
        (b) Mirrors \& Frequency based OCSs
    \end{tabular}
    \caption{An example of an RDCN. (a) A topology with eight ToR switches and two optical spine switches. Each switch is configured with a matching between its ports (violet arcs) to create optical circuits between ToR pairs. (b) Examples of two types of reconfigurable optical switches that enable \emph{dynamic} topologies through reconfiguration of the matching between ports.}
    \label{fig:OCS}
\end{figure}

\section{Paradigm Shift: Reconfigurable Topologies and Its Enabling Optical Technologies}\label{sec:reconfigurable}

\begin{quotation}
    \textit{“It's really more recently that we're performing real-time, what I call 'topology engineering', to match communication patterns.”} 
    \par \hfill--Amin Vahdat (Google), Interviews, Section \ref{sec:Conclusion}
\end{quotation}

With the maturation of optical switching technologies, RDCNs have emerged as an innovative and attractive alternative to traditional static topology datacenter network designs. Such datacenter networks based on dynamic topologies represent a paradigm shift, standing in stark contrast to the fixed nature of state-of-the-art datacenter network topologies. In particular, reconfigurable topologies can reduce the high 
bandwidth tax of static networks, by dynamically providing shorter paths or even direct connections. 

RDCNs are enabled by optical switching technologies, which facilitate quick changes in topological interconnections, e.g., between datacenter racks. More specifically, RDCNs are often realized using OCSs: rather than actually rewiring the datacenter, the (electrical) ToRs are connected to a set of OCSs, and at any given time, each OCS provides a set of physical connections (circuits) between racks, via suitably configured matching between its ports. Figure 3 (a) shows an example of such a setup with eight ToRs and two optical switches, each with a matching configuration between its ports that creates optical circuits between two ToRs.
 
The revolutionary idea underlying RDCN design is that each switch’s matching (i.e., circuits) can be dynamically reconfigured,, by remapping the input ports to different output ports. In turn, this reconfiguration changes the physical network topology. Various technologies enable such matching reconfiguration in optical switches. For example, a topology can be reconfigured using OCSs based on micro-electromechanical systems (MEMSs)~\cite{poutievski2022jupiter}. In a MEMS-based OCS, the optical signals are forwarded via mirrors, which can be actuated and tilted to steer each signal to the desired port (see Figure~\ref{fig:OCS} (b) top).
In an alternative technology, each source ToR switch can change the frequency (wavelength, color) at which its signal is transmitted, and passive gratings are used in the OCS to direct, every entering frequency/color to a different output port~\cite{sirius} (see Figure~\ref{fig:OCS} (c) bottom).
Gratings are very simple and passive building blocks with no moving parts and do not need any power. Each grating simply diffracts or routes incoming light to an output port based on its wavelength. This concept can hence be used for physical-layer switching by equipping each node’s transceivers with tunable lasers: the wavelength (represented by color in the figure) serves as a proxy for the destination address.

There are a variety of alternative switching technologies enabling reconfigurable datacenters~\cite{hall2021survey}, including nanosecond tunable lasers~\cite{lange2020sub-nanosecond}, digital micromirror devices~\cite{ghobadi2016projector}, and liquid crystal on silicon wavelength selective switches~\cite{shakeri2017traffic}. Different technologies have different advantages, in terms of performance or high-precision manufacturing requirements. For example, liquid- crystal-on- silicon systems can typically be packed more compactly due to the absence of moving parts. While conventional MEMS-based beam-steering cross-connects cannot provide sub-millisecond switching functionalities for large datacenters, alternative noncrossbar selector switch architectures can offer improved performance and scalability~\cite{rotornet}. 
Even faster OCS can be realized using microcombs and semiconductor optical amplifiers~\cite{sirius}. 

OCSs generally support very high link bandwidths at low per-bit cost and power because they simply redirect light from one port to another, independent of the data rate and without requiring header parsing for its operation. In this sense, each optical circuit is similar to a wired connection. Expensive optical–-electrical–-optical conversions are not necessary at optical switches;: therefore, in an OCS-based datacenter, far fewer such conversions are required (e.g., are limited only at ToRs), hence removing a bottleneck\footnote{Note that electrical switches are usually also connected by fiber-optic links.}.
Furthermore, such an optical infrastructure can remain unchanged even as the electrical infrastructure is upgraded incrementally and heterogeneously upgraded.

Different types of reconfigurable switches correspond to different types of RDCNs, as we will discuss in more detail in the next section. However, independent of the technology used, all such dynamic networks share a common advantage and a common disadvantage. By providing topological shortcuts, dynamic datacenter networks can reduce the bandwidth tax:, namely, the overhead incurred in multi-hop forwarding. 
For example, while the network diameter in Figure \ref{fig:biregular} is four, in Figure \ref{fig:OCS}(a), the \emph{expected route length} can be as low as one, depending on the demand and the optimized topology.
However, dynamical optical switching also comes at a price. In contrast to static networks, RDCNs additionally incur a \emph{latency tax}; optical links are not available during reconfiguration, and reconfiguration takes time and therefore entails overhead (e.g., related to buffering and throughput \cite{griner2021cerberus, addanki2023mars,amir2022optimal} ). While the latency tax is inherent to dynamic topologies, its magnitude depends on the technology and architecture. For example, if the demand changes can be predicted or are known, the latency tax can be reduced by saving real-time computations; but in any case, it can not be completely eliminated.

\begin{figure}[t]
    \centering
    \begin{tabular}{>{\centering}m{0.33\textwidth}>{\centering}m{0.33\textwidth} >{\centering}m{0.33\textwidth}}
    \multicolumn{3}{c}{\includegraphics[width=\textwidth]{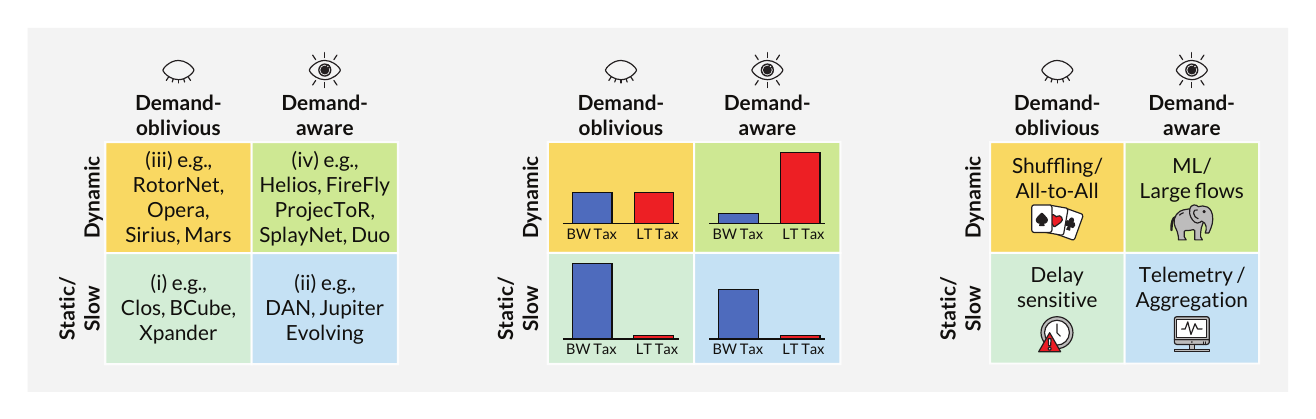}} \\
    (a) Systems  & (b) Taxes & (c) Traffic
    \end{tabular}
    \caption{(a) Classification of the design space along two dimensions, i) static vs. dynamic topologies and ii) demand-oblivious vs. demand-aware topologies, corresponding to four topology types. (b) Each topology type incurs different levels of bandwidth (BW) and latency (LT) taxes. (c) Different types of traffic are better served by different types of systems.}
    \label{fig:spectrom}
\end{figure}

\section{The Spectrum of Topology Engineering Solutions}\label{sec:challenges}

\begin{quotation}
    \textit{“Demand oblivious doesn't necessarily preclude either centralized or decentralized control. However, the history of all networking is, you know, that centralized control only works to a certain level.”} 
    \par \hfill--George Papen (UCSD), Interviews, Section \ref{sec:Conclusion}
\end{quotation}

Optical switching technologies herald a new era of RDCNs. In fact, they enable an entire spectrum of datacenter architectures and topology engineering solutions, the extent and implications of which researchers are still trying to understand.

A fundamental dimension along which possible datacenter network designs can be classified, concerns whether they are \emph{static or dynamic}. As discussed above, traditional networks are static in nature and rely on multi-hop forwarding (thus incurring a bandwidth tax), while emerging dynamic topologies can provide more direct connectivity but incur a latency tax (which depends on the frequency of topology reconfiguration). Moreover, there is a second, and, as we will see, \emph{independent} dimension, along which datacenter networks can be classified: whether they are \emph{demand-oblivious} or \emph{demand-aware}.
We hence propose to classify datacenter designs in two dimensions: static vs. dynamic and demand-oblivious vs. demand-aware.
This classification implies \emph{four} main topology types, as illustrated in Figure \ref{fig:spectrom} (a).
For each of these four types, multiple realizations are possible.

\begin{enumerate}[(i)]
\item {\bf Static, demand-oblivious:} Traditional datacenter networks, as discussed in Section \ref{sec:traditional}, are based on \emph{static} and \emph{demand-oblivious} topologies, e.g., fat-tree or expander graphs topologies~\cite{xpander,clos,jupiter}. 
Recall that these networks are optimized for the worst-case traffic scenario by establishing many disjoint short paths.  
\item {\bf Static, demand-aware:} Instead of the traditional worst-case optimization, static networks may also be optimized in a demand-aware manner, toward a specific and fixed workload. In this case, we assume that the demand matrix is known (or predicted) in advance.  For example, Google \cite{poutievski2022jupiter} recently introduced a practical and efficient slowly evolving (on the scale of days) topology for RDCNs based on demand prediction.
\item {\bf Dynamic, demand-oblivious:} Researchers started exploring highly \emph{dynamic}, but \emph{demand-oblivious} topologies, e.g., relying on rotor (or round-robin) spine switches that \emph{periodically} and quickly (on the scale of nanoseconds) reconfigure the topology in a predefined and fixed  manner~\cite{rotornet,opera,sirius,addanki2023mars,amir2022optimal}. 
\item {\bf Dynamic, demand-aware:} Dynamic and demand-aware topologies, also called \emph{self-adjusting networks}, 
can be reconfigured according to the current traffic pattern. Examples include Helios~\cite{helios1}, Firefly~\cite{firefly}, ProjecToR~\cite{ghobadi2016projector}, and, recently Cerberus~\cite{griner2021cerberus}, among others
~\cite{zhou2012mirror,kandula2009flyways,chen2014osa,ghobadi2016projector,teh2020flexspander,eclipse,splaynet}.
\end{enumerate}

How do these designs fare against each other? It turns out that the optimal architecture depends on the traffic. Traffic with a significant spatial structure (e.g., a small number of elephant flows) can benefit from demand-aware network designs, while uniform traffic may be better served by demand-oblivious networks. In contrast, only dynamic and demand-aware (self-adjusting) networks can optimally adapt to the temporal structure of evolving traffic. To systematically understand the tradeoffs, one should recall that the throughput achievable by static topologies is typically limited by the bandwidth tax, whereas dynamic topologies are subject to a latency tax. This latency tax is likely to be higher in demand-aware networks, which, in addition to physical reconfiguration, also require a more complex control plane (e.g., to learn about the traffic and/or optimize for it). 
We illustrate this in Figure \ref{fig:spectrom} (b). In summary, static topologies have a lower latency tax (LT Tax) than dynamic topologies, while demand-oblivious topologies have a higher bandwidth tax (BW Tax) than demand-aware topologies.

A major cause of suboptimal performance in datacenter networks is the \emph{mismatch} between some common traffic patterns and the network topologies that serve them, as presented in Figure \ref{fig:spectrom} (c). In particular, different optical topologies on the design spectrum provide different tradeoffs. For example, mice flows, or delay-sensitive packets should be served on a static topology; sending them on dynamic topologies that require time for reconfiguration may result in high flow completion times and violation of their latency constraints. On the other hand, elephant flows (e.g., in ML~\cite{khani2021sip} or datamining applications~\cite{greenberg2009vl2}), will benefit from dynamic demand-aware topologies. Since the reconfiguration time is relatively short compared to the transmission time for such large flows, the reconfiguration time cost can be amortized, and the throughput can be improved by establishing direct links between frequently communicating pairs. To give one more example, it has been shown that due to its all-to-all nature, shuffling traffic, e.g., for map-reduce or all-gather collective operations in ML, can benefit from dynamic demand-oblivious topologies that provide periodic direct connectivity between all rack pairs~\cite{rotornet,opera,sirius}.   

Given these tradeoffs and the desire to match different traffic types to their optimal network infrastructures, many emerging datacenter network designs are hybrid designs, supporting multiple modes of operation. We will discuss some of them in more detail in the next section.

\section{Examples of Reconfigurable Datacenter Networks Designs}\label{sec:examples}

Reconfigurable datacenters have developed significantly in recent years, from early prototypes to initial deployments. To highlight this evolution and the spectrum of related system designs, we present some  examples in more detail in the following.

RotorNet~\cite{rotornet} and Sirius~\cite{sirius} are examples of demand-oblivious, periodically reconfigurable networks. RotorNet pioneered the idea of decoupling switch reconfiguration from traffic patterns. Each switch simply and independently rotates through a fixed, static set of configurations that provide uniform bandwidth between all endpoints. This approach allows for a fully decentralized control plane: such round-robin switch scheduling does not require demand estimation, schedule computation, or schedule distribution, yet delivers full bisection bandwidth for uniform traffic. To support arbitrary traffic patterns, RotorNet relies on a form of Valiant load balancing. RotorNet can accommodate a hybrid optical/electronic architecture to extend support to traffic with lower latency requirements. RotorNet has also inspired several additional demand-oblivious RDCN designs, such as Opera~\cite{opera} (essentially a sequence of expander graphs), Microsoft's Sirius~\cite{sirius} (providing unprecedented nanoseconds-granularity reconfiguration
in an optically-switched network), and Mars~\cite{addanki2023mars}.
To achieve very fast reconfigurations, Sirius uses a combination of tunable lasers and simple, passive gratings that route light signals based on their wavelengths. Sirius also comes with novel congestion-control and time-synchronization mechanisms.
 
Google Jupiter~\cite{poutievski2022jupiter} is an example of a demand-aware, slowly evolving network, supporting incremental heterogeneity and dynamic application adaptation.  Its OCSs dynamically map an optical fiber input port to an output port through two sets of MEMS mirrors that can be rotated in two dimensions to create arbitrary port-to-port mappings. Enabled by this OCS layer, Google has eliminated the spine layer from its datacenter networks, instead connecting heterogeneous aggregation blocks in a direct mesh, and therefore moving beyond fat-tree topologies. 

ProjecToR~\cite{ghobadi2016projector}  was the first proposal of a dynamic demand-aware datacenter network developed by Microsoft Research based on free-space optics. This network design provides a high scalability and fan-out, enabled by dense micromirror devices (DMDs). The DMDs are placed near the ToRs and paired with mirrors fixed to the ceiling above the racks. Each DMD is programmed to target a specific mirror at the ceiling, guiding light to another ToR in the datacenter. In ProjecToR, possible links are divided into two categories: dedicated and opportunistic. Dedicated links carry small flows, possibly over multiple hops, and their configurations change at coarse time scales (e.g., daily). Opportunistic links carry large flows over single hops, and their configurations change rapidly based on current demand. 

Cerberus~\cite{griner2021cerberus}
is motivated by the observation that each existing network design, be it static or dynamic, demand-oblivious or demand-aware, has its own specific advantages. In particular, different optical topologies provide different tradeoffs, and the design used should depend on the demand.
Accordingly, Cerberus is a hybrid datacenter network design that maximizes throughput by serving each traffic class using the datacenter topology that best matches its requirements. Some optical switches are operated in a demand-aware manner, creating a demand-aware topology that can best serve elephant flows; some optical switches are demand-oblivious, creating a demand-oblivious dynamic topology that is best suited to all-to-all traffic; and yet other optical switches create a static topology that can best serve delay-sensitive short flows. 

\section{Models of Reconfigurable Datacenter Networks}\label{sec:model}

For studying and evaluating reconfigurable networks,
evolving graphs~\cite{michail2016introduction} are a natural abstract model. In contrast to the original evolving graph model, a critical issue in our model is the \emph{reconfiguration time} consumed for edge changes. In turn, this reconfiguration time introduces latency and can influence important metrics such as the throughput. Recently, Mars~\cite{addanki2023mars} introduced a formal definition of periodic evolving graphs in the context of oblivious reconfigurable networks, enabling the study of fundamental performance tradeoffs in such networks. The model 
can in fact universally capture static, dynamic, demand-oblivious, and demand-aware topologies. 
Formally, we define evolving graphs:

\begin{definition}[Evolving Graph]\label{def:evolving-graph}
    An evolving graph denoted by $\mathcal{G}=(V,\mathcal{E})$ is a sequence of directed graphs, in which one graph is defined for each timeslot $t\in \mathbb{W}$ (the set of whole numbers). The directed graph at time $t\in \mathbb{W}$ is defined as $G_t = (V, E_t)$, where $V$ is the set of $n$ vertices and $E_t\subseteq V\times V$ is the set of directed edges at time $t$. Note that an edge $e \in V\times V$ may appear in multiple edge sets $E_t$ at different times $t$. 
    The timeslot length is denoted by $\Delta$ [sec].
    If not stated otherwise the capacity of edge $e \in E_t$ at time $t$, denoted by $c_t(e)$, is $c$ for all edges and all times. 
\end{definition}

Evolving graphs can be restricted to specific families of graphs, for example, where for every $t$, $G_t$ is a degree-$d$ bounded graph or a $r$-regular graph, or where for every $t$,  the edge set $E_t$ is periodic with period $\Gamma$ i.e.,\ $E_{t+\Gamma}=E_{t}$ ($\Gamma$-periodic graphs).

To capture the reconfiguration time for establishing a new edge, we set the timeslot length $\Delta$ equal to the reconfiguration time. Additionally, we assume that when a new edge $e$ is established at time $t$, i.e.,\ $e \in E_t$ and $e \notin E_{t-1}$, $e$ cannot be used at time $t$, i.e.,\ $c_t(e) = 0$, since this timeslot is devoted to reconfiguration. 
Note that before reconfiguration, designers will need to consider graceful packet drain, or packet loss, or rerouting via different paths \cite{zerwas2023duo}.
Additional restrictions may also apply, for example, an \emph{inter-reconfiguration} time $c' \cdot \Delta$, ensuring that reconfigurations cannot occur too close to each other in time, e.g., since some computation or data collection is needed. 

\begin{figure}[t]
    \centering
    \begin{tabular}{>{\centering}m{0.4\textwidth}>{\centering}m{0.3\textwidth} >{\centering}m{0.3\textwidth}}
    \multicolumn{3}{c}{\includegraphics[width=\textwidth]{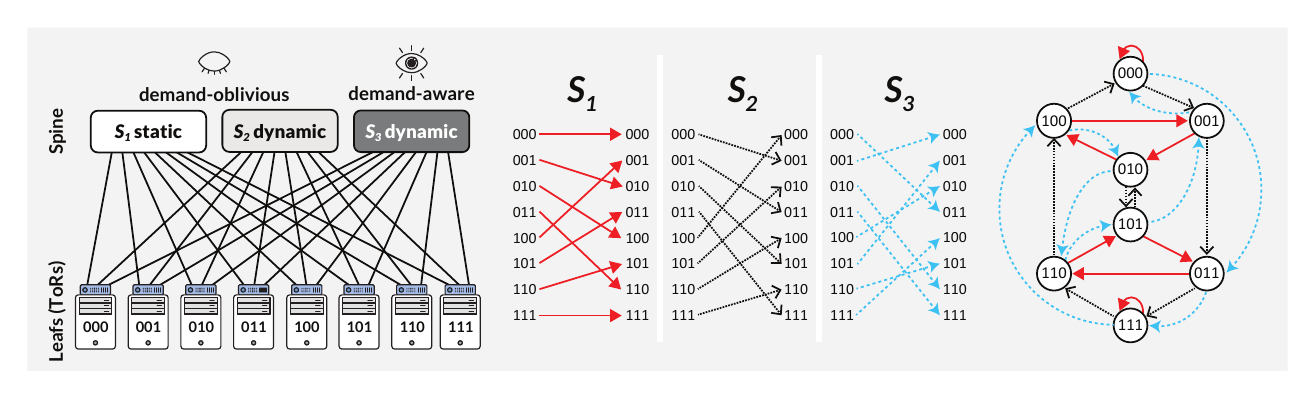}} \\
    (a) TMT architecture & (b) Switch matching & (c) $G_t=(V,E_t)$
    \end{tabular}
    \caption{: A 3-regular evolving graph with eight nodes implemented in a ToR–matching–ToR (TMT) network example with eight ToR switches and three optical spine switches: one has a static matching configuration, and the other two have dynamic matching configurations (one demand-oblivious and one demand-aware). Each spine switch has eight input and output ports. (a) TMT architecture. (b) The matching configurations inside each of the three switches at time $t$. (c) The resulting 3-regular topology at time $t$, $G_t$.
    }
    \label{fig:deisgn:tmt_network}
\end{figure}

Evolving graphs can also be realized on the basis of a standard datacenter  architecture that connects ToRs.
This ToR-Matching-ToR (TMT) topology model, which is common in the literature~\cite{rotornet,sirius,opera,griner2021cerberus}, involves $n$ ToR (leaf) switches that are interconnected by a set of $k$ optical spine switches. Each spine switch provides a $n \times n$ \emph{directed matching} between its input and output ports. Depending on the type of spine switch (or its mode of operation), the matching within each spine switch can be static or can dynamically change over time, in either a demand-oblivious or a demand-aware manner. The current topology $G_t$ at time $t$, is implemented as the union of the $k$ matching configurations. Figure~\ref{fig:deisgn:tmt_network} (a) presents an example of a TMT model with eight ToR switches $000, 001, \dots 111$, and three optical spine switches $S_1, S_2, S_3$ that use different matchings scheduling. $S_1$ uses static matching, $S_2$ uses demand-oblivious dynamic matching, and $S_3$ uses demand-aware dynamic matching.
Each ToR-spine link in the figure represents one directed uplink (to the spine switch's input port) and one directed downlink (from the spine switch's output port).
Figure~\ref{fig:deisgn:tmt_network} (b) shows the matching configurations used within each spine switch at time $t$. These matching configurations can overlap (increasing the capacity of a ToR-to-ToR connection) or be disjoint. Dynamic switches can change their matching configurations over time. Figure~\ref{fig:deisgn:tmt_network} (c) illustrates the topology of the evolving graph at time $t$, $G_t$. The set of nodes $V$ is the ToRs switches, and the set of edges $E_t$ is the union of the edges in the three matchings. Using these matchings some special topologies can be created. For example, careful readers may observe that the matchings of $S_1$ and $S_2$ create a 2-regular De Brujin graph that has some nice properties, like supporting greedy routing; it was recently used in Duo~\cite{zerwas2023duo} and Mars~\cite{addanki2023mars}.
 It is important to note that although we abstractly consider $k$ spine switches here, 
each implementing an $n \times n$ matching, each such matching configuration can be split across a set of several smaller switches in practice, as in Sirius~\cite{sirius}.

\section{Conclusion, Challenges Ahead and Interviews with Experts}\label{sec:Conclusion}

While reconfigurable datacenter topologies are an exciting vision, it is also bold. While the first concepts and specific technologies for reconfigurable datacenter topologies are being actively developed, e.g., at Google, the field is very young. Early deployments are small-scale and reconfigure slowly, however, there are developments and efforts to support more fine-granular adaptions and hence further improve network performance. Reconfigurable networks, both demand-oblivious and demand-aware, introduce additional complexities, but also bear the potential to simplify certain aspects of network operations. 
In general, the transition to highly dynamic topologies will likely be long and strenuous. 

\paragraph{\bf Operational and deployment challenges}
In terms of operational and deployment challenges, we currently do not have a good understanding of the implications of dynamic networks from the perspective of the other layers of the networking stack. When is it beneficial or even needed to adapt, e.g., the routing layer or congestion control mechanisms for reconfigurable datacenter networks (see, e.g.~\cite{li2024uniform,zerwas2023duo,mukerjee2020adapting,liang2024negotiator} for early works in these directions)? Should we maintain a fixed network layering structure, or is a cross-layer approach justified, e.g., for joint topology optimization and traffic engineering? What are the implications of reconfigurations on, for example, buffer management~\cite{addanki2023mars} or the traffic pattern itself? Which functionality can be supported in software (which is more flexible),  and which in hardware (which is more complex but more performant)? Where to deploy the optical circuit switches --- is the typical assumption that one layer of OCSs is enough in RDCNs justified? Should optical switches connect datacenter racks or rather entire pods?
What is the optimal interplay between electrical packet switches and optical circuit switches in future hybrid datacenters?
And what is the cost of such architectures? 
Etc.

To answer some of the above questions our article comes with complementary video interviews in which three leading experts share with us their perspectives on reconfigurable datacenter networks, the state of the art, and the challenges ahead.

\paragraph{\bf Video interviews with experts}  
Manya Ghobadi (professor at MIT) shares with us her insights on the special structure of datacenter traffic compared to other communication traffic, and how this structure can be exploited in reconfigurable networks such as ProjecToR (\href{https://youtu.be/tN7jJdqTwHk}{link to interview});
George Papen (professor at UCSD) describes the main technologies that enable optical circuit switches that are used in reconfigurable datacenter networks such as RotorNet, and how different technologies fare against each other (\href{https://youtu.be/fdJurYxTY08}{link to interview}); 
Amin Vahdat (Vice President at Google) discusses the advantages and disadvantages of the different datacenter architectures, and why machine learning applications may particularly benefit from reconfigurable networks (\href{https://youtu.be/IxcV1gu8ETA}{link to interview}). 

A key challenge, according to Amin Vahdat, concerns the control system of the reconfigurable datacenter network, and its scalability.
He envisions that future, improved control systems may enable a faster and more fine-granular adaption of the topology, which can further improve datacenter network performance. 
If decisions have to be taken within seconds or even milliseconds, it can be difficult or impossible to collect sufficient information about the network traffic in real-time, and generally, the control system may have to combine both decentralized components (for fast reaction and robustness) and centralized components (for global visibility and optimization). Multiple controllers may also be organized in a hierarchical manner. Existing distributed system software, by default, is not used to exert control at such granularity. Despite these challenges, Vahdat emphasizes that reconfigurable datacenter networks also bear a potential for \emph{simplifying} datacenter planning and operations in many aspects, e.g., by naturally supporting heterogeneous network elements and therefore incremental hardware upgrades.

A big opportunity but also challenge for reconfigurable networks comes especially from machine learning applications, according to Manya Ghobadi. Machine learning workloads often feature significant temporal and spatial structure, and traffic is fairly predictable, which can simplify the control plane of reconfigurable datacenters. However, collecting and characterizing such traffic patterns and tailoring a reconfigurable network for a very specific workload, such as distributed ML, still imposes significant operational and research challenges.

According to George Papen, dynamic topologies also introduce novel failure scenarios, for which current debugging tools are insufficient or not applicable at all. Since in large networks, failures can happen frequently, detection and reaction must be fast. 
Failure scenarios can also be technology-dependent and
generally, different technologies currently come with different tradeoffs. Which architecture will be most attractive in terms of the performance and robustness it provides, as well as its hardware and operational costs, is currently unclear and will likely evolve over time.
At least in terms of energy costs, optical switches are expected to be more efficient compared to traditional switches~\cite{sirius}. 

\paragraph{\bf Research challenges} Last but not least, the vision of reconfigurable datacenter networks 
and topology engineering also poses interesting graph-theoretical research problems. 
The design of self-adjusting networks is related to graph spanners and embedding problems \cite{opodis24dan,avin2017demand}, as well as to self-adjusting data structures \cite{ccr18san}.
In \cite{avin2017demand}, we presented initial theoretical results, relating the average path length in a demand-aware network to the entropy of the demand. 
However, for many performance metrics, including throughput \cite{addanki2023mars,amir2022optimal}, we currently only know very little about how they can be optimized algorithmically in datacenter networks with controlled topologies. 

\section*{Acknowledgments}

We would like to thank our interviewees Manya, Amin, and George, and all our co-authors working with us on self-adjusting networks over the last years. Research supported by the European Research Council (ERC), grant agreement no.~864228 (AdjustNet), as well as by the Israeli Science Foundation, grant ISF~2497/23.

\bibliographystyle{abbrv}
\bibliography{cacm}

\end{document}